\documentclass[manuscript]{aastex}
\usepackage{natbib}
\usepackage{subfigure}
\newcommand{\revision}[1]{#1}
\title{Magnetohydrodynamics of the Weakly Ionized Solar Photosphere}

\author{Mark C. M. Cheung\affil{Lockheed Martin Solar and Astrophysics Laboratory \\ 3251 Hanover St, Palo Alto, CA 94304, USA; cheung@lmsal.com}}
\author{Robert H. Cameron \affil{Max-Planck-Institut f{\"u}r Sonnensystemforschung \\ Max-Planck-Str. 2, D-37191 Katlenburg-Lindau, Germany}}

\begin{abstract}

We investigate the importance of ambipolar diffusion and Hall currents for high-resolution comprehensive (`realistic') 
photospheric simulations. To do so we extended the radiative magnetohydrodynamics code~\emph{MURaM} to use the generalized Ohm's law under the assumption of local thermodynamic equilibrium. We present test cases comparing analytical solutions with numerical simulations for validation of the code. Furthermore, we carried out a number of numerical experiments to investigate the impact of these neutral-ion effects in the photosphere. We find that, at the spatial resolutions currently used (5-20~km per grid point), the Hall currents and ambipolar diffusion begin to become significant -- with flows of 100 m/s in sunspot light bridges, and changes of a few percent in the thermodynamic structure of quiet-Sun magnetic features. The magnitude of the effects is expected to increase rapidly as smaller-scale variations are resolved by the simulations. \\
\end{abstract}

\begin{document}
\section{Introduction}
In recent years, three-dimensional numerical magnetohydrodynamics (MHD) simulations have been a boon to our 
understanding of a variety of phenomena in the solar atmosphere and underlying convection zone. By including an increasingly extensive array of important physical effects – e.g. radiative transfer, magneto-convection, heat conduction along magnetic field lines, time-dependent hydrogen ionization etc – such models are beginning to unravel many mysteries of solar magnetism. To varying degrees, the most important physical effects have been captured by state-of-the-art radiative MHD simulations~\citep{Voegler:MURaM,SteinNordlund:SmallScaleMagnetoconvection,Hansteen:DynamicFibrils,Steiner:HorizontalFields,Gudiksen:Bifrost}, which aim to model the solar atmosphere as realistically as computationally feasible. The ability of these models to reproduce the observational properties for a wide variety of phenomena, such as solar faculae~\citep{Keller:Faculae,Carlsson:SolarMagnetoconvection,DePontieu:Faculae}, emerging flux  regions~\citep{SteinNordlund:SmallScaleMagnetoconvection,Cheung:FluxEmergenceInGranularConvection,Cheung:SolarSurfaceEmergingFluxRegions,MartinezSykora:TwistedFluxEmergence, 
MartinezSykora:TwistedFluxEmergenceII,TortosaAndreu:FluxEmergence,Stein:SolarFluxEmergenceSimulations,Cheung:ARFormation}, pores~\citep{Cameron:Pores, Kitiashvili:Spontaneous},  
sunspots~\citep{Rempel:PenumbralStructure,Rempel:SunspotStructure,Cheung:ARFormation,Stein:ProtoARs} and 
spicules~\citep{Hansteen:DynamicFibrils,MartinezSykora:SpiculelikeStructures,MartinezSykora:TypeIISpicules} etc is encouraging. 

However it is also clear that the magnetic diffusivities and viscosities used in the simulations are too high: the molecular values correspond to length scales much smaller than the presently achievable resolution. In these circumstances the simulations have been moving towards ever higher resolution~\citep{Voegler:SolarSurfaceDynamo,PietarilaGraham:TurbulentMagneticFields,Moll:Universality}. 

Indeed the simulations have reached the point where, at some locations in the atmosphere, we are approaching some of the lengthscales
associated with the microphysical processes. In particular both the Hall current and ambipolar diffusion begin to influence the plasma dynamics in the photosphere at resolutions of 5-20~km and on timescales of seconds to minutes. The Hall effect results from the relative immobility of ions (relative to electrons) to react sufficiently quickly to the Lorentz force. Associated with this Hall drift of ions with respect to electrons is an electric field parallel to the Lorentz force. 
This effect is important for dynamical changes in the plasma at frequencies between the ion and electron gyrofrequencies ($\omega_j= eB/m_jc$).  For a fully ionized plasma and a magnetic field in the kiloGauss range, the proton gyrofrequency is $\omega_p \sim 10^7$ Hz and the Hall effect thus can be safely neglected. The solar photosphere, however, is only weakly ionized  and the gyrofrequency quoted above does not apply. In a weakly ionized plasma, most of the inertia is in neutrals, which do not respond directly to the Lorentz force. The neutrals do however feel the magnetic field indirectly through collisions with the ions. Since the photosphere is collisionally dominated,  the neutrals effectively give the ions extra inertia, and the Hall effect becomes 
important at much lower frequencies. The relevant frequency becomes $\omega_i^* = eB/m_i^*c$, where $m_i^*=(\varrho_i+\varrho_n)/n_i$ is the effective ion mass. For the low values of ionization fraction in the 
photosphere and lower chromosphere, the effective proton gyrofrequency (or Hall frequency) can be as small as 
$10$ Hz~\citep{PandeyWardle:HallMHD,PandeyVranjesKrishnan:WavesInSolarPhotosphere}.

In addition to the Hall effect, ambipolar diffusion can have an important influence on the dynamics of the photosphere and chromosphere. Since neutrals do not directly experience the Lorentz force, they have the ability to drift across magnetic field lines. Like classical Ohmic diffusion associated with a scalar magnetic resistivity, ambipolar diffusion is accompanied by the dissipation of an electric current (the Pedersen current), which converts magnetic energy into heat. Unlike Ohmic (associated with finite Spitzer conductivity) and ambipolar diffusion, the Hall effect is not dissipative and hence does not directly contribute to plasma heating~\citep{Priest:SolarMHD,Parker:Conversations}. Nevertheless, the Hall effect does redirect magnetic energy throughout the system (i.e., it contributes a term to the Poynting flux) and thus potentially has important indirect influences on wave damping and plasma heating.

Although state-of-the-art 3D radiative MHD simulations have so far neglected the Hall effect and ambipolar diffusion, there is a considerable body of theoretical and observational work that point to the importance of neutral-ion interactions in the solar atmosphere. For instance, the effect of Pedersen current dissipation on wave damping and chromospheric heating have been investigated by~\citet{DePontieuHaerendel:WeaklyDampedAlfvenWaves},~\citet{Goodman:2000,Goodman:2004,Goodman:2005,Goodman:2010},~\citet{Khodachenko:WaveDamping},~\citet{Leake:WaveDamping},~\citet{PandeyVranjesKrishnan:WavesInSolarPhotosphere},~\citet{Singh:AlfvenWavesInPartiallyIonizedAtmosphere} and~\citet{Khomenko:HeatingOfChromosphere}. 
All these studies conclude that neutral-ion collisions (i.e. Pedersen contribution to Joule heating) is the dominant dissipation mechanism in the chromosphere (as opposed to a single-fluid MHD model with only Ohmic dissipation from Spitzer resistivity). Furthermore, the ensuing chromospheric heating may be the driver of spicules~\citep{DePontieuHaerendel:WeaklyDampedAlfvenWaves, James:SpiculesI, James:SpiculesII}. 
Sakai and co-authors have used two-fluid (neutral hydrogen and proton components) MHD simulations to model the collision of current-carrying loops in the chromosphere~\citep{SakaiTsuchimotoSokolov:TwoFluidLoops} as well as to model penumbral microjets~\citep{SakaiSmith:PenumbralMicrojets}. In the former study, they found substantial differences in the results between single-fluid MHD and corresponding two-fluid MHD simulations, 
from which they conclude that the single-fluid MHD approximation (for fully-ionized plasmas) is inappropriate for the description of chromospheric dynamics. Additionally, they note that their single-fluid MHD simulation neglects Pedersen dissipation (i.e. ambipolar diffusion), which should be included to capture the relevant neutral- ion effects.

\citet{Leake:EmergenceThroughPartiallyIonizedAtmosphere} and~\citet{Arber:EmergenceThroughPartiallyIonizedAtmosphere} performed 
idealized numerical simulations of magnetic flux emergence into the solar atmosphere. Using a 1d height-dependent profile for the hydrogen ionization fraction consistent with the VAL-C quiet-Sun model~\citep{VAL:1981} they performed flux emergence simulations with and without ambipolar diffusion. A comparison between simulations with and without ambipolar diffusion showed marked differences. First of all, ambipolar diffusion preferentially dissipates the Pedersen current. Since the Pedersen current is perpendicular to the magnetic field, its dissipation leaves the field reaching the corona in a force-free state. Furthermore, the departure from the frozen-in condition of the magnetic field allows it to traverse the chromospheric layers with substantially less mass uplift (i.e. less work against gravity) than in simulations without ambipolar diffusion.

While most of these theoretical studies have concentrated on the chromosphere, the effects are also important in the upper photosphere. In light of the potential importance of the Hall effect and ambipolar diffusion on solar atmospheric dynamics, the purpose of the present study is to investigate their influence on a number of physical scenarios in the solar photosphere.  The paper is structured as follows: Section~\ref{sec:equations} describes model equations and implementation of the numerical code. Section~\ref{sec:test_cases} presents results from tests cases used for validation of the code as well as in obtaining physical insight into the basic properties of the effects.  We then study the effect of ambipolar diffusion on the structure of a quiet-Sun magnetic flux sheet and the effect of Hall currents in a sunspot light bridge. Our conclusion is that we have indeed reached the point where their effects are becoming dynamically important.

\section{Equations and implementation}
\label{sec:equations}
The radiation MHD code~\emph{MURaM}~\citep{Voegler:MURaM,Rempel:PenumbralStructure}~already takes into account important physical effects such as radiative energy exchange in the energy equation as well as partial ionization in the equation of state. Both are important for the energy balance and stratification of the convection zone and overlying photosphere. Here we extend the model to incorporate the Hall effect and ambipolar diffusion. In comparison to the MHD equations for fully-ionized plasmas, this extension involves two additional terms in generalized Ohm's law~\citep{Parker:Conversations,PandeyWardle:HallMHD}. The modified induction equation is

\begin{eqnarray}
\frac{\partial \vec{B}}{\partial t} & =& \nabla \times [ \vec{v}\times\vec{B} - \frac{4\pi\eta}{c}\vec{j}  -\frac{1}{e n_e}\vec{j}\times\vec{B}    
  + (\vec{j}\times\vec{B}) \times \vec{B} \frac{D^2}{c \varrho_i\nu_{in}} ], \label{eqn:GeneralizedInduction}
\end{eqnarray}
\noindent where $\vec{B}$ is the magnetic field, $\varrho_i$ and $\varrho_n$ the ion and neutral mass densities, respectively, $D = \varrho_n/(\varrho_i+\varrho_n)$ is the neutral mass fraction, $\vec{v} = D\vec{v}_n + (1-D)\vec{v}_i$ is the bulk velocity of the center of mass of the neutral and ionized components, $\vec{j}= c (4\pi)^{-1} \nabla\times\vec{B}$ is the current density, $\eta$ is the Ohmic diffusivity, $c$ the speed of light, $e$ the electron charge, $n_e$ the electron number density, and $\nu_{in}$ the rate of ion-neutral collisions. The third and fourth terms on the right-hand side represent Hall effect and ambipolar diffusion, respectively.

Eq. (\ref{eqn:GeneralizedInduction}) can be written in the form
\begin{equation}
\frac{\partial \vec{B}}{\partial t} = \nabla \times [\vec{u} \times \vec{B} - \frac{4\pi}{c}\eta \vec{j}],\label{eqn:SimplifiedInduction}
\end{equation}
\noindent where
\begin{eqnarray}
\vec{u} &=& \vec{v} + \vec{v}_{\rm Hall} + \vec{v}_{\rm Amb},\label{eqn:GeneralizedVelocity}
\\
\vec{v}_{\rm Hall} & = & - H \nabla\times \vec{B}, \label{eqn:HallVel}\\
\vec{v}_{\rm Amb} & = & M  (\nabla \times \vec{B}) \times \vec{B}, \label{eqn:AmbVel}\\
H & =  &\frac{c}{4\pi e n_e},\label{eqn:hall_coeff}\\
M & = & \frac{D^2}{4\pi \varrho_i\nu_{in}}. \label{eqn:ambipolar_coeff}
\end{eqnarray}
\noindent Here, $\vec{u}$ is the effective velocity operating on $\vec{B}$. In the presence of the Hall effect and ambipolar diffusion, departures of $\vec{u}$ from $\vec{v}$ means that magnetic field lines are no longer frozen into the plasma. This consequence, however, is not equivalent to a change in the field topology. Since the two effects act like velocities advecting, compressing and stretching the field, the field topology is in fact preserved. Another way to look at this is to consider the electric fields associated with the two effects. Both the Hall and ambipolar electric fields are perpendicular to $\vec{B}$, which means that they do not lead to magnetic reconnection. Ohmic diffusion must act in order for reconnection to take place. 

The~\emph{MURaM} code solves the MHD equations in conservation form, which means that physical quantities on the grid are updated based on the net fluxes of the quantities into and out of individual grid cells. This implementation ensures that the total mass, energy, and momentum within the simulation domain are conserved in the absence of net fluxes through the domain boundaries. For studying how the Hall and ambipolar terms change the energy budget in the solar atmosphere, we implemented the effects in such a way such that the numerical scheme remains conservative. 

By virtue of Eqs. (\ref{eqn:SimplifiedInduction}) and (\ref{eqn:GeneralizedVelocity}), the generalized induction equation can be written in conservation form as follows:
\begin{equation}
\frac{\partial \vec{B}}{\partial t} + \nabla \cdot [\vec{u} \otimes \vec{B} - \vec{B} \otimes \vec{u}] = -\nabla \times (\eta \nabla \times \vec{B}), 
\end{equation}
\noindent where $\vec{u} \otimes \vec{B}$ and $\vec{B} \otimes \vec{u}$ denote tensor products. The energy equation must also be modified. In conservation form, it reads

\begin{eqnarray}
\frac{\partial e}{\partial t} &+&\nabla \cdot \left[ \vec{v}(\frac{1}{2}\varrho v^2 + \varrho \varepsilon + p) +\vec{u}\frac{B^2}{4\pi}- \frac{1}{4\pi}\vec{B}(\vec{u}\cdot\vec{B})\right] \nonumber \\
& = & \frac{1}{4\pi}\nabla \cdot (\vec{B}\times\eta\nabla\times\vec{B})+ \nabla\cdot (\vec{v}\cdot\underline{\underline{\tau}}) \nonumber \\
& &+ \nabla\cdot(K\nabla T)+ \varrho(\vec{g}\cdot\vec{v}) + Q_{\rm rad}, \label{eqn:HallMHDEnergy}
\end{eqnarray}
\noindent where $e = \varrho\varepsilon + \frac{1}{2}\varrho v^2 + B^2/8\pi$ is the total energy density, $\underline{\underline{\tau}}$ the viscous tensor, $\varepsilon$ is the specific internal energy density, $\vec{g}$ the gravitational acceleration ($\vec{g}=2.74\times 10^4$ cm s$^{-2}$ at the solar surface), $K$ the thermal conductivity, and $Q_{\rm rad}$ the radiative heating/cooling term, which is determined by means of solving the radiative transfer equations along rays in various directions~\citep{Voegler:MURaM}. The gas pressure $p$ and temperature $T$ are functions of $\varrho$ and $\varepsilon$ and are determined from a look-up table during runtime. In the absence of the Hall and ambipolar effects (i.e. $\vec{u} = \vec{v}$), Eq. (\ref{eqn:HallMHDEnergy}) reduces to the energy equation solved in the original version of~\emph{MURaM}. The momentum equation remains unchanged since terms that are quadratic in the drift velocity $\vec{v}_D = \vec{v}_i-\vec{v}_n$ (i.e. the difference in the bulk velocities of the ion and neutral species) can be safely neglected under photospheric conditions~\citep{HasanSchuessler:HeatingByDownflows,PandeyWardle:HallMHD}.

\emph{MURaM} uses 4-th order Runge-Kutta explicit time integration. The inclusion of the Hall and ambipolar terms in the induction and energy equations introduces new constraints on the CFL stability condition~\citep*{CFL}. Since the Hall and ambipolar effects express themselves in the equations as effective velocities $v_{\rm}$, the associated time-step constraints imposed by the CFL condition are $\Delta t_{\rm Hall} = \Delta x/|\vec{v}_{\rm Hall}|$ and $\Delta t_{\rm Amb} = \Delta x/|\vec{v}_{\rm Amb}|$ where $\Delta x$ is the grid-spacing. Since $\vec{v}_{\rm Hall}$ and $\vec{v}_{\rm Amb}$ are first-order in $j \sim \Delta B/\Delta x$, both $\Delta t_{\rm Hall}$ and $\Delta t_{\rm Amb}$ are in fact quadratic in $\Delta x$. For this reason, both the Hall and ambipolar effects can be considered as non-linear diffusion terms in the induction equation.

\subsection{Coefficients in the Hall and ambipolar terms}
The coefficients $H = c(4\pi e n_e)^{-1}$ and $M=D^2/c\varrho_i\nu_{in}$ in Eq. (\ref{eqn:hall_coeff}) and (\ref{eqn:ambipolar_coeff}) 
are precalculated onto a look-up table for use during runtime. There are several assumptions made in creating the tables. The first is that the plasma is in local thermal equilibrium. The second assumption is the relative fraction of each element in the plasma. The Saha-Eggert equations can then be solved to obtain the number densities of each chemical species and the number density of free electrons. We use the code described in \cite{Wittmann:Saha}. The coefficient in Eq. (\ref{eqn:hall_coeff}) can be determined without further assumptions. The coefficient in Eq. (\ref{eqn:ambipolar_coeff}) requires an additional approximation to obtain $\nu_{in}$. Three different sources are given in \cite{DePontieu:Coll}; we chose to use choice 'a' which corresponds to that given by \cite{Osterbrock:Coll}.

In the chromosphere, the plasma is sufficiently tenuous such that LTE is no longer a suitable assumption for determining ionization fractions of hydrogen~\citep{Kneer:TimeDependentIonization}. Therefore, a time-dependent approach to calculating number densities of the different hydrogen specifies must be used~\citep{Leennarts:Hion}. This is not yet implemented in the code. However, the numerical experiments presented below focus on dynamics at the photospheric layers.

\section{Test cases}
\label{sec:test_cases}

Here we describe two simulation setups used as test cases for validation of our implementation of the Hall and ambipolar diffusion effects.

\subsection{Influence of the Hall effect on plane-polarized Alfv\'en waves}\label{sec:PlanePolarizedAlfvenWaves}
Consider the one-dimensional problem of Alfv\'en waves propagating along the $x$-direction in the presence of a guide field, $B_x$. The transverse ($B_y $ and $B_z$) components of the field are assumed to be small compared to the guide field $B_x$, so that second-order terms in these quantities can be neglected. For a pure Alfv\'en wave, the total background mass density ($\varrho_0$) is constant and uniform. For this test example we neglect the effect of ambipolar diffusion. The evolution equations are then
\begin{eqnarray}
\varrho_0 \frac{\partial \vec{v}}{\partial t}& =& \frac{1}{4\pi}{(\nabla \times \vec{B}})\times \vec{B}, {\rm~and} \\
\frac{\partial \vec{B}}{\partial t} &=& \nabla \times \left[\vec{v}\times\vec{B} - H(\nabla\times\vec{B})\times \vec{B} \right].
\end{eqnarray}

\noindent \revision{Let the solutions for $\vec{v}$ and $\vec{B}$ be of the form }
\[ \left( \begin{array}{c}
v_x \\ v_y \\ v_z 
\end{array} \right) =  \left( \begin{array}{c}

0 \\
v' e^{i(\omega t - kx)} [\cos{\sigma t} - i\delta\sin{\sigma t}] \\
v' e^{i(\omega t - kx)} [\sin{\sigma t} + i\delta\cos{\sigma t}]
\end{array} \right), \]

\[ \left( \begin{array}{c}
B_x \\ B_y \\ B_z 
\end{array} \right) =  \left( \begin{array}{c}
b_0 \\
b' e^{i(\omega t - kx)} \cos{\sigma t} \\
b' e^{i(\omega t - kx)} \sin{\sigma t}
\end{array} \right),\]

\noindent \revision{where $\sigma$ and $\delta$ are still to be determined. $v'$ and $b'$ are amplitudes for the traveling wave solutions and are assumed constant (for a chosen $k$). The $y$-component of the momentum equation reduces to}
\begin{eqnarray}
v'[i\omega\cos{\sigma t} - \sigma\sin{\sigma t} + \delta \omega \sin{\sigma t}- i\delta\sigma\cos{\sigma t}]&=&\frac{-ikb_0}{4\pi\varrho_0}b'\cos{\sigma t}, \label{eqn:vp}
\end{eqnarray}
\noindent \revision{Since $v'$ and $b'$ are constants, equating coefficients of $\sin{\sigma t}$ in Eq. (\ref{eqn:vp}) gives $\sigma = \delta \omega$. Substituting this back into Eq. (\ref{eqn:vp}) and eliminating the common factor $\cos{\sigma t}$, we obtain}
\begin{equation}
v' = \frac{-\omega}{\omega^2-\sigma^2} \frac{kb_0}{4\pi\varrho_0}b'
\end{equation}

\noindent \revision{Now the $y$-component of the induction equation gives}
\begin{eqnarray}
b'[i\omega\cos{\sigma t}-\sigma\sin{\sigma t}] = -ikb_0v'[\cos{\sigma t} - i\delta\sin{\sigma t}] - b'Hb_0k^2\sin{\sigma t}
\label{eqn:bp}
\end{eqnarray}
\noindent \revision{Using Eq. (\ref{eqn:vp}) to substitute for $v'$ in (\ref{eqn:bp}), we obtain}
\begin{equation}
i\omega \cos{\sigma t} - \sigma\sin{\sigma t} = i \left(\frac{b_0^2k^2}{4\pi\varrho_0 }\right)\left(\frac{\omega}{\omega^2-\sigma^2}\right)[\cos{\sigma t}-i\delta\sin{\sigma t}] - Hb_0k^2\sin{\sigma t}
\label{eqn:complexdispersion}
\end{equation}
\noindent \revision{Equating coefficients for $\cos{\sigma t}$ in Eq. (\ref{eqn:complexdispersion}), we obtain the dispersion relation
}
\begin{equation}
\omega^2 = v_A^2k^2 + \sigma^2,\label{eqn:dispersion}
\end{equation}
\noindent \revision{where $v_A=b_0(4\pi\varrho_0)^{-1/2}$ is the Alfv\'en speed. Similarly, equating coefficients for $\sin{\sigma t}$ in Eq. (\ref{eqn:complexdispersion}) and using the dispersion relation (\ref{eqn:dispersion}) gives}
\begin{equation}
\sigma = \frac{Hb_0k^2}{2} = \frac{c b_0 }{8\pi n_e e }k^2.\label{eqn:precession}
\end{equation}
\noindent \revision{The dispersion relation Eq. (\ref{eqn:dispersion}) is similar to that of a pure Alfv\'en wave except for a correction term. For the phase speed $\omega/k$, the correction term is of order $O(\vert \vec{v}_{\rm Hall} \vert)$, where $\vec{v}_{\rm Hall}$ is the Hall velocity as defined in Eq. (\ref{eqn:HallVel}).} The physical interpretation of the above result is as follows. Suppose the Alfv\'en wave of wavenumber $k$ is initially plane-polarized in the $y$-direction. Under the action of the Hall effect, the plane of polarization of this wave would precess about the $x-$axis at a rate equal to $\sigma$.

\begin{figure}
\centering
\includegraphics[width=0.48\textwidth]{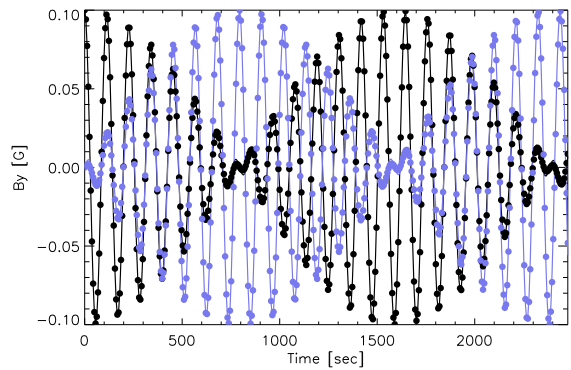}
\label{fig:test1}
\caption{The two magnetic field components $B_y$ (black) and $B_z$ (purple) as functions of time (sampled at $x=0$) from the
numerical (dots) and analytic (solid line) solutions for $k=2\pi \times 10^{-2}$ km$^{-1}$. The fast time variation corresponds to the Alfv\'en waves,
the slow variation to a rotation of the plane of polarization due to the Hall effect.}\label{fig:hall_plane1}
\end{figure}
\begin{figure}
\centering
\includegraphics[width=0.48\textwidth]{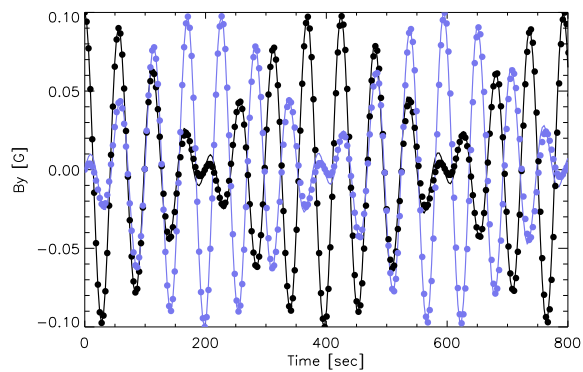}
\caption{Same as Fig.~\ref{fig:hall_plane1} but with twice the wavenumber. Note that the time interval shown in this plot is shorter than that shown in Fig.~\ref{fig:hall_plane1}.}\label{fig:hall_plane2}
\end{figure}

For the following one-dimensional test case, we choose to use physical parameters that are typical of photospheric conditions. The $x$-component of the magnetic field is uniform with $B_x(t=0) = b_0 = 100$ G. Superposed on this uniform field is a sinusoidal profile of the tranverse field $B_y(x,t=0) = b'\cos(k x)$, where $b'= 0.1$ G and $k = 2\pi\times 10^{-2}$ km$^{-1}$. The one-dimensional simulation domain has width $L_x = 100$ km so the domain is spanned by exactly one wavelength of the transverse perturbation in $B_y$. The initial total pressure of the unperturbed magnetic field and plasma is $p_{\rm tot} = p_{\rm gas } + B^2/8\pi = 4 \times 10^4$ dyn cm$^{-2}$. The mass density is approximately $\varrho=10^{-7}$ g cm$^{-3}$ and the temperature is $T=6000$ K. Gravity is switched off so that the only force initially acting on the plasma is magnetic tension. For the sake of carrying a test simulation, we impose a constant Hall coefficient of $H=10^8$ cm$^2$ s$^{-1}$ G$^{-1}$ and switch off the ambipolar term (i.e. $M=0$). The grid spacing used for the simulation is $\Delta x = 1$ km and periodic boundary conditions are used.

The black and purple dots in Fig.~\ref{fig:hall_plane1}, respectively, show values of $B_y$ and $B_z$ at $x=0$ sampled at different times during the simulation. The black and purple curves in the same plot show the corresponding analytical solution to the problem. Since the initial condition consists of the standing wave, the solution consists of a superposition of modes propagating in the positive and negative $x$-directions, viz.
\begin{eqnarray}
B_y(x,t) &=& \frac{b'}{2}\cos(\sigma t)[\cos(kx-\omega t) \nonumber \\
&&-\cos(-kx-\omega t)],\\ 
&=& b' \cos(\sigma t) \cos(kx) \cos(\omega t), \\
B_z(x,t) &=& b'\sin(\sigma t) \cos(kx) \cos(\omega t).
\end{eqnarray}
\noindent For the set of physical parameters given above, the rate of precession of the plane of polarization is $\sigma = 2\times 10^{-3}$ Hz. This corresponds to a rotation period of $\tau = 2\pi\sigma^{-1}= 3.2\times 10^3$ s. The intrinsic oscillation frequency of the Alfv\'en wave is $\omega = 5.5\times 10^{-2}$ Hz. For the same initial setup, but with twice the wavenumber (i.e. $k=4\pi\times 10^{-2}$ km), $\omega =0.11$ Hz and $\sigma = 8\times 10^{-2}$ Hz. Fig.~\ref{fig:hall_plane2} shows the solution for this case. Compared to the case when $k=2\pi\times 10^{-2}$ km$^{-1}$ (Fig.~\ref{fig:hall_plane1}), we see here that for $k=4\pi\times 10^{-2}$ km$^{-1}$ there are half as many Alfv\'enic oscillations for each complete rotation of the plane of polarization since $\sigma \propto k^2$ and $\omega \propto k$.

\subsection{Collapse of current layers due to ambipolar diffusion}
\label{sec:ambipolar_test}
\revision{An effect of ambipolar diffusion is the reduction of the width of
current sheets~\citep{Parker:Flares,BrandenburgZweibel:AmbipolarDiffusion}}. In the absence of Ohmic diffusion, the time-independent solution for the transverse magnetic field in the neighborhood of the tangential discontinuity has the profile
\begin{equation}
B \propto x^{1/3},\label{eqn:brandenburg_zweibel}
\end{equation}
\noindent where $x$ is the coordinate in the direction normal to the curernt sheet. They also point out that the current density at the neutral line increases without bound if Ohmic diffusion is absent (namely, the constant of proportionality in Eq.(\ref{eqn:brandenburg_zweibel}) increases with time). Acting alone, ambipolar diffusion does not lead to magnetic reconnection. This property is a direct consequence of the fact that the ambipolar term in the induction iquation can be written as an effective velocity $\vec{v}_{\rm Amb}$ acting on $\vec{B}$. So ambipolar diffusion merely acts to sharpen the current layers which in turn allows Ohmic diffusion to occur at an enhanced rate. 

\begin{figure}
\centering
\includegraphics[width=0.48\textwidth]{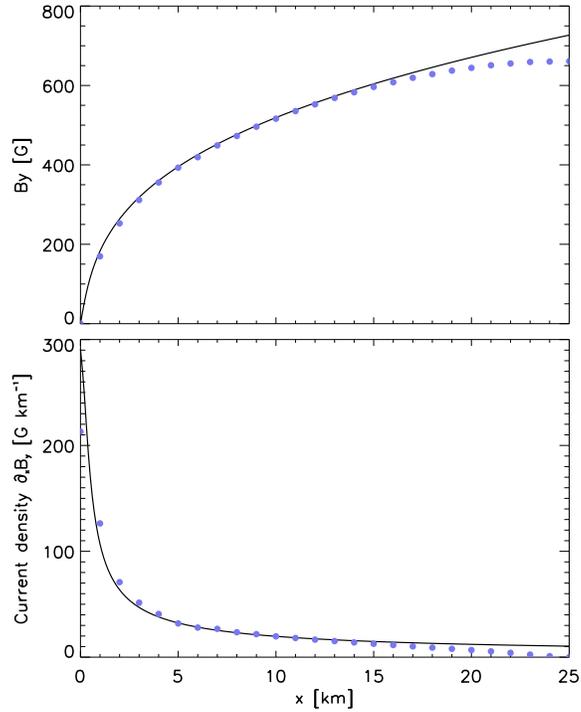}\label{fig:ambipolar_test}
\caption{The magnetic field (upper panel) and current density (lower panel) distributions for the test case corresponding to a collapsed current sheet in which Ohmic and ambipolar diffusion balanced each other. In both panels, the solid line indicates the semi-analytical solution and the purple dots indicate the solution from the simulation. The deviation of the two away from the current sheet is expected.}
\end{figure}

To test our implementation of ambipolar diffusion, we consider a one-dimensional system with the following initial distribution:
\begin{equation}
\vec{B}(x,t=0) = (0,B_0 \sin(kx),0).\label{eqn:sinusoidal_by}
\end{equation} 
\noindent We set $B_0=1000$ G, $k=2\pi \times 10^{-2}$ km$^{-1}$ and $x\in[0,100]$ km. The plasma has an initial uniform temperature of $T=6000$ K and the system is in total pressure equilibrium, with  $p_{\rm tot} = p_{\rm gas } + B^2/8\pi = 4.4 \times 10^4$ dyn cm$^{-2}$. For this simulation, we set a constant Ohmic diffusivity of $\eta=10^8$ cm$^2$ s$^{-1}$, a constant ambipolar coefficient of $M=5\times 10^3$ cm$^2$ s$^{-1}$ G$^{-2}$, and switch off the Hall term (i.e. $H=0$). Periodic  boundary conditions at $x=0$ and $x=100$ km were imposed.

\revision{We follow the analysis of~\citet{Parker:Flares} to seek a time-independent solution for the collapsed current sheet~\emph{including Ohmic diffusion}}. For the one-dimensional setup described above, the induction equation reduces to 
\begin{equation}
\frac{\partial B_y}{\partial t} = \frac{\partial }{ \partial x} \left[ (M B_y^2 + \eta)\frac{\partial B_y}{\partial x}\right]. \label{eqn:ambipolar_test_induction}
\end{equation}
\noindent For a time-independent solution and for constant $M$ and $\eta$, Eq. (\ref{eqn:ambipolar_test_induction}) can be integrated to yield 
\begin{equation}
\frac{M}{3}B_y^3 + \eta B_y = Cx + D,\label{eqn:ode}
\end{equation}
\noindent where $C$ and $D$ are constants determined by boundary conditions. For our setup here, $B_y$ is antisymmetric about $x=0$. This means $B_y(x=0)=0$, which implies $D=0$. Since the simulation setup employs periodic boundary conditions at $x=0$ and at $x=100$ km, there is not an infinite reservoir of magnetic flux which can be transported into the diffusion region for reconnection. As such, the simulted system never really reaches a true steady state and the integration constant $C$ depends on how much flux has reconnected from the initial state.

In the absence of Ohmic diffusion ($\eta=0$), Eq. (\ref{eqn:ode}) can be directly solved to yield the relation given by Eq. (\ref{eqn:brandenburg_zweibel}). Otherwise, we identify that for a specific spatial location $x$ (distance from the neutral line), Eq. (\ref{eqn:ode}) is a cubic polynomial in $B_y$. To determine for $B_y(x)$ we used a numerical solver for the polynomial corresponding to various values of $x$. This is plotted as solid lines in both panels of Fig.~\ref{fig:ambipolar_test}. The corresponding solution from the simulation is plotted as purple dots. For $C=2.85\times 10^5$ G cm s$^{-1}$, the simulated and semi-analytical solutions closely resemble each other in the vicinity of the current sheet. For reasons already discussed, they are expected to deviate far away from the current sheet.

\section{Simulation results}
 In the following sections we present simulation results on how the additional effects impact the physical evolution of magnetic field in conditions mimicking the solar atmosphere.

\subsection{Magnetic reconnection in the presence of ambipolar diffusion and radiative cooling}
In Section~\ref{sec:ambipolar_test}, we examined how ambipolar diffusion steepens gradients in the magnetic field to sustain current sheets. For the sake of code validation, we assumed a constant ambipolar coefficient $M$. However, this assumption neglects an important negative feedback loop, which proceeds in the following fashion. Both ambipolar diffusion and Ohmic dissipation act to heat plasma in and around the non-ideal region. The heating of the plasma leads to an increase in the ionization degree of the plasma, which in turn could suppress the efficiency of ambipolar diffusion (i.e. decrease $M$). So although ambipolar diffusion may act to steepen up a current sheet, it does not necessarily result in runaway magnetic reconnection. 

\begin{table*}
\centering
\begin{tabular}{c|c|c|c}
\hline
Run & Ambipolar coeff. $M$ &  Cooling & Flux cancelled by $t=820$ s\\
\hline
\texttt{ReA} & $2.3\times 10^4$ cm$^2$ s$^{-1}$ G$^{-2}$ & No & $28\%$\\
\texttt{ReB} & Variable & No & $24\%$\\
\texttt{ReC} & Variable & Yes & $51\%$\\
\hline
\end{tabular}\caption{Set of reconnection experiments with and without variable ambipolar diffision and/or Newton cooling. Eq. (\ref{eqn:ambipolar_coeff}) is used for evaluation of the variable ambipolar diffusion $M$.}\label{table:reconnection}
\end{table*}

To investigate this effect, we carried out three one-dimensional numerical experiments (see Table ~\ref{table:reconnection}). The initial condition in all cases consists of a current sheet with 
\begin{eqnarray}
B_y &=& 1000 {\rm~G}, {\rm~if~}x\ge0,\nonumber\\
& = & -1000 {\rm~G} {\rm~otherwise}.
\end{eqnarray}
\noindent The other components of $\vec{B}$ are zero. The plasma initially has uniform temperature of $T=5000$ K and the uniform total pressure $p_{\rm tot} = p_{\rm gas } + B^2/8\pi = 4.13 \times 10^4$ dyn cm$^{-2}$ (except at the tangential discontinuity). This choice leads to $\beta = 8\pi p_{\rm gas}/B^2 = 0.04$, so that dynamics are dominated by magnetic forces. The domain spans $x\in[-300,300]$ km and the grid-spacing is $\Delta x=6$ km. Periodic boundary conditions are imposed at $x=\pm 300$ km.

\begin{figure}
\centering
\includegraphics[width=0.45\textwidth]{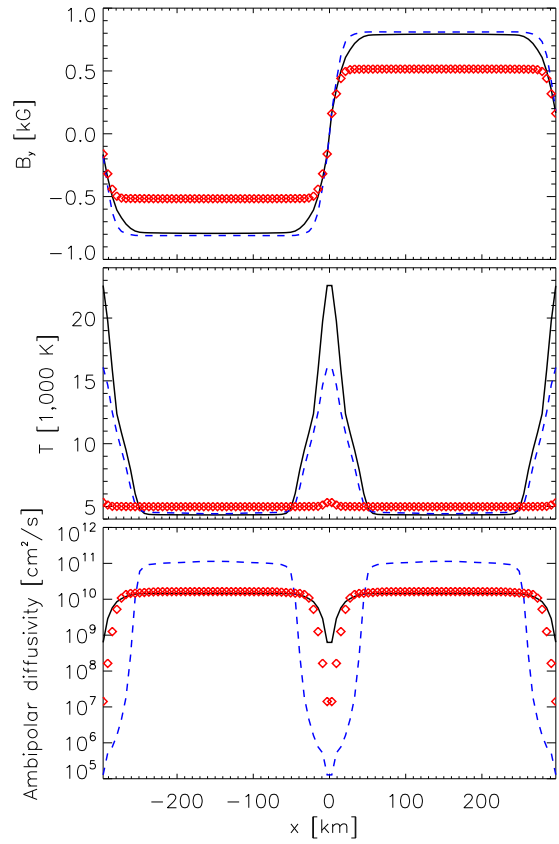}
\caption{The distribution of magnetic field, temperature and ambipolar diffusivity $\eta_{\rm Amb}$ in simulation runs~\texttt{ReA} (solid line),~\texttt{ReB} (dashed line) and~\texttt{ReC} (diamonds) at time $t=820$ s.}\label{fig:reconnection_runs_compare}
\end{figure}

For the initial thermodynamic state of the plasma, Eq. (\ref{eqn:ambipolar_coeff}) gives $M=2.3\times 10^4$ cm$^2$ s$^{-1}$ G$^{-2}$. In run~\texttt{ReA}, we keep $M$ uniform and constant with this value. In run~\texttt{ReB}, we allow $M$ to have values consistent with the local thermodynamic state at any time by using our look-up table. The radiative cooling term in the energy equation is switched off in both runs~\texttt{ReA} and~\texttt{ReB}. In run~\texttt{ReC}, we introduce a cooling term in the energy equation to mimic the energy loss from the plasma as it heats up above the temperture of its surroudings.~\emph{MURaM} is capable of calculating this source/sink term self-consistently by solving the radiative transfer equation along multiple directions, but this is not appropriate for a one-dimensional setup. The cooling term in run~\texttt{ReC} is of the form

\begin{equation}
Q_{\rm rad} = -a\varrho[T^4-T_0^4],
\end{equation}
\noindent where $\varrho$ is the local mass density, $T$ the local plasma temperature, $T_0=5000$ K and $a = 10^{-5}$ erg s$^{-1}$ g$^{-1}$ K$^{-4}$. $T>T_0$ implies $Q_{\rm rad} < 0$ and the plasma loses energy. This decreases the plasma temperature (and ionization fraction) and provides a way for the ambipolar diffusion to continue acting. 

Figure~\ref{fig:reconnection_runs_compare} plots the distribution of $B_y$, $T$ and the ambipolar diffusivity $\eta_{\rm Amb} = M B^2$ for all three simulation runs at time $t=820$ s. Between $t=0$ and this instant, the fraction of the initial unsigned flux that has cancelled in runs~\texttt{ReA},~\texttt{ReB} and~\texttt{ReC} is $28\%$, $24\%$ and $51\%$, respectively. The reason more flux has cancelled in run~\texttt{ReA} than in run~\texttt{ReB} (a difference which increases with time) is that in the latter, the temperature in the current sheet due to Ohmic and ambipolar dissipation suppresses $M$ and thus $\eta_{\rm Amb}$.~\revision{This negative feedback mechanism is absent in run~\texttt{ReA} since the ambipolar coefficient is set to be constant.} In run~\texttt{ReC}, radiative cooling in the current sheet maintains the action of ambipolar diffusion and leads to a sustained flux cancellation rate of almost double that of the two other runs.

\subsection{Vertical flux sheet}

\begin{figure*}
\centering
\includegraphics[width=0.63\textwidth]{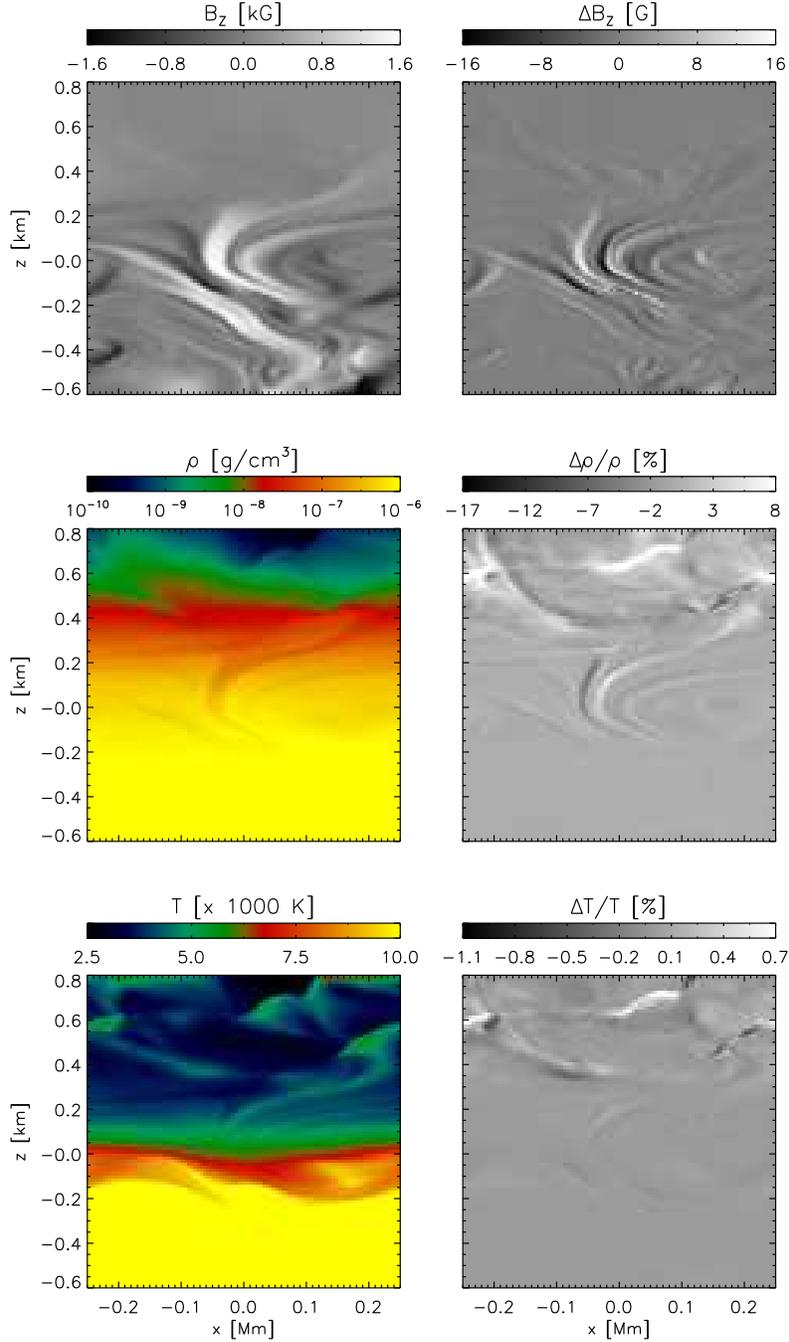}
\caption{Effect of ambipolar diffusion on the evolution of a small magnetic flux sheet in the photosphere. The left panels show snapshots of the vertical component of the magnetic field (top), the mass density (middle) and the gas temperature (bottom) for the run with ambipolar diffusion switched off. The right panels show the difference between the snapshot and a similar snapshot where ambipolar diffusion was included. 
These snapshots are taken after 2.5 minutes of evolution from a common initial condition.}\label{fig:tube}
\end{figure*}

We have also investigated the effect of ambipolar diffusion on the structure of small-scale magnetic features in the photosphere. For this purpose we set up a two dimensional simulation extending vertically from about 700~km below to a height of $700$ km above the photospheric base ($\tau_{\mathrm Ross}=1$). The horizontal extent of the box was chosen to be 500~km and contains one convective roll. This was done so that only one flux sheet is ever present which simplifies the analysis. A grid spacing of 5~km was used in the horizontal direction and 14~km in the vertical direction. Non-magnetic convection was allowed to develop for several hours, after which an initially uniform vertical field of 100 G was imposed, and the system was evolved for several additional hours. This state was then used as an initial condition for runs with and without ambipolar diffusion. Fig.~\ref{fig:tube} shows the temperature, density and  vertical component of the magnetic field after 2.5 minutes for the case without ambipolar diffusion and the difference resulting from the inclusion of ambipolar diffusion. The effect is relatively weak (a few percent) consistent with 
expectations \citep[e.g.][]{PandeyWardle:HallMHD}.

\subsection{Magnetoconvection in a sunspot umbra}

In this simulation, we examine the influence of the Hall effect on magnetoconvection at field strengths found within sunspot umbrae and in the vicinity of sunspot light bridges. In this two-dimensional simulation, we begin with statistically relaxed hydrodynamical convection driven by radiative cooling at the photospheric base. The computational domain spans $18.4$ Mm in the horizontal direction and $6.1$ Mm in the vertical direction with horizontal and vertical grid spacings of \revision{$9$} km and \revision{$6$} km respectively. After introducing a uniform vertical magnetic field with $B_z = 1500$ G, we ran the simulation for approximately one hour to allow the magnetoconvection to relax. The result is the development of strong magnetic regions with deep Wilson depressions interspersed with upflow regions with relatively weak field. This type of cross-sectional structure may represent umbral dots~\citep{Schuessler:UmbralConvection} and sunspot light bridges~\citep{Cheung:ARFormation}. 

\begin{figure*}
\centering
\includegraphics[width=\textwidth]{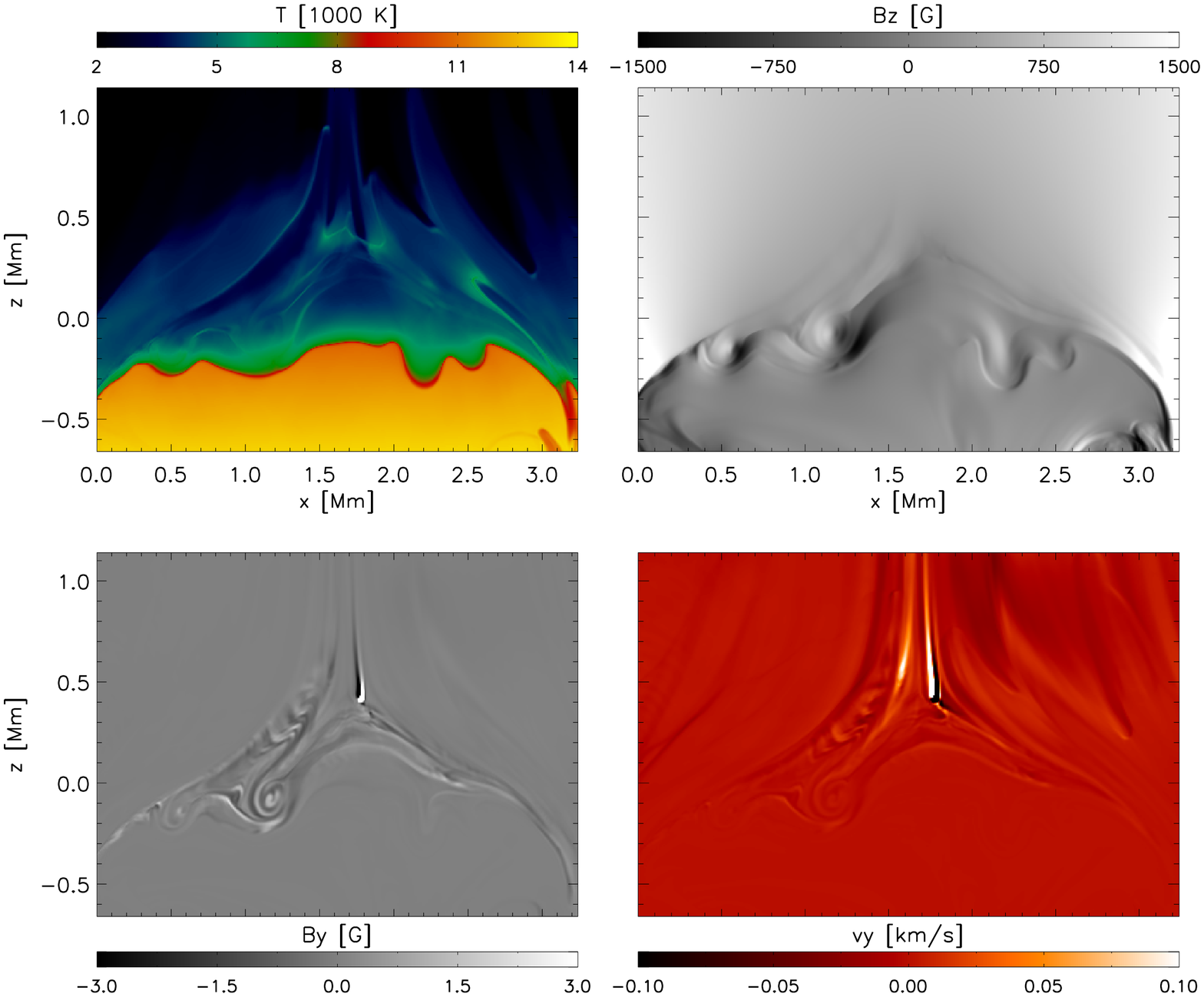}
\caption{Cross-sections of physical quantities from a 2D simulation of umbral magnetoconvection with treatment of the Hall effect. The Hall term leads to an out-of-plane component of the magnetic field ($B_y$, lower left panel). The presence of this third component leads to a Lorentz force which acts to accelerate plasma in the out-of-plane direction ($v_y$, lower right panel).}~\label{fig:hall_fluxrolls}
\end{figure*}

Figure~\ref{fig:hall_fluxrolls} shows the distribution of physical quantities of a structure in the simulation resembling a sunspot light bridge. In both the temperature and vertical magnetic field panels, one finds the characteristic cusp shape where canopy fields from adjacent umbral regions merge~\citep{Schuessler:UmbralConvection}. In the absence of the Hall effect, the magnetic and velocity fields are confined to the $x$- and $z$-components. The sharp interfaces between weak-field regions within the light bridge and the strong field regions outside it acts as the site of high current density, which in the presence of the Hall term leads to an effective $\vec{v}_{\rm Hall}$ in the $y$-direction.~\revision{This component of the effective velocity in the $y$-direction is non-uniform and} leads to the generation of a $y$-component (out-of-plane) of the magnetic field. After $B_y$ is generated by the Hall term, convective flows advect the flux from the layer and back into the convection zone. These flux rolls can be seen in the bottom left panel of Fig~~\ref{fig:hall_fluxrolls}.

We emphasize that the strength of $B_y$ is only $\sim 5$ G, which is much weaker than the kilogauss fields in the umbra. Thus in the current simulation, the fields generated by the Hall currrent do not strongly influence the dynamical evolution of the system. Nevertheless, the presence of non-zero $B_y$ leads to a Lorentz force which accelerates an actual (i.e. not just $\vec{v}_{\rm Hall}$) out-of-plane flow $v_y$ (see bottom right panel of Fig~\ref{fig:hall_fluxrolls}) to speeds of up to $0.1$ km/s. Such flows are completely absent in a simulation without the Hall effect.

\section{Discussion}
Both ambipolar diffusion and the Hall current have now been included in a code designed to produce 'realistic' 
simulations of the solar photosphere. The implementation has been validated, and the physical effects such as the rotation of the plane of polarization of Alfv\'en waves and the steepening of current sheets has been reproduced.

In simulations of magnetic reconnection with ambipolar diffusion, we find that ambipolar diffusion acts to sharpen current layers. This is a well-known result from the study by~\citet*{BrandenburgZweibel:AmbipolarDiffusion}. However, our study also shows that the temperature increase associated with ambipolar and Ohmic dissipation in the current sheet rapidly suppresses ambipolar diffusion. In this sense, ambipolar diffusion is a self-limiting mechanism for enhancing magnetic reconnection. Only when we include an energy loss mechanism for the plasma (in this case a radiative cooling term) does the effect remain important. This result has important implications for studies of chromospheric dynamics. The effect of ambipolar diffusion in the chromosphere may be severely overestimated when the ionization fractions and energy loss mechanisms in chromospheric plasma are not treated in a realistic fashion~\citep[e.g. such as time-dependent H ionization, see][]{Leennarts:Hion}.

In a two-dimensional simulation of umbral magnetoconvection, we examined how the Hall effect generates a third (out-of-plane) component of the magnetic and velocity fields at interface layers between weakly magnetized light bridges and neighboring strong field umbral regions. At the grid spacing of \revision{$\sim 10$} km, the Hall term results in generation of the third component of the magnetic field at the level of $\sim 5$ G from kilogauss umbral fields. The generation of the out-of-plane component of the magnetic field faciliates a magnetic tension force which accelerates plasma out of the plane to speeds of up to $0.1$ km/s. 

At the current grid resolution, the influence on dynamics of the system due to the presence of the Hall effect begins to become important. Furthermore, we expect that the relative importance of both the Hall and ambipolar effects to increase with grid resolution. This is due to the expectation that with decreasing grid spacing $\Delta x$, numerical simulations would be able to resolve increasing sharp current layers (since $j\sim\Delta B / \Delta x$). Since both the Hall and ambipolar velocities ($\vec{v}_{\rm Hall}$ and $\vec{v}_{\rm Amb}$, respectively) in the induction equation are linear in $\vec{j}$, their amplitudes relative to the actual fluid velocity $\vec{v}$ and to numerical diffusion should increase linearly with grid resolution. A Hall-induced velocity of $0.1$ km s$^{-1}$ at $\Delta x = 20$ km, when scaled to $\Delta x = 2$ km, already yields a velocity of $1$ km s$^{-1}$. Thus the linear trend is unlikely to continue down to Ohmic dissipation scales ($1 \lesssim \Delta x_{\rm Ohm} \lesssim 10^3$ m in the photosphere). With this in mind, future investigations will examine how Hall amd ambipolar effects will, at sufficiently high spatial resolution, come to be important for dynamics at small scales.

\acknowledgements
We are grateful to Manfred Sch\"ussler for interesting and informative discussions about the research presented in this paper. MCMC acknowledges support from NASA contract NNX10AC02G. Numerical simulations presented here were made possible by NASA's High-End Computing Program. The simulations presented in this paper were carried out on the Pleiades cluster at the Ames Research Center. We thank the Advanced Supercomputing Division staff for their technical support.

\bibliographystyle{apj}
\bibliography{references,apj-jour}
\end{document}